# DBSCAN-based Vehicle Clustering and UAV Placement for NOMA-based Resource Management in Cellular V2X Communications

Hossein Davoudi[1], Behrouz Shahgholi Ghahfarokhi[1], Neda Moghim[1,2], and Sachin Shetty[2]
[1]Department of Computer Engineering, University of Isfahan, Isfahan, Iran
{hdavoudi, shahgholi}.eng.ui.ac.ir
[2]Center for Secure & Intelligent Critical Systems, Old Dominion University, VA, USA
{ n.moghim@eng.ui.ac.ir, sshetty@odu.edu}

*Abstract*— In the future wireless networks, terrestrial, aerial, space, and maritime wireless networks are integrated into a unified network to meet the needs of a fully connected global network. Nowadays, vehicular communication has become one of the challenging applications of wireless networks. In this article, we aim to address the radio resource management in Cellular V2X (C-V2X) networks using Unmanned Aerial Vehicles (UAV) and Non-orthogonal multiple access (NOMA). The goal of this problem is to maximize the spectral efficiency of vehicular users in Cellular Vehicle-to-Everything (C-V2X) networks under a fronthaul constraint. To solve this problem, a two-stage approach is utilized. In the first stage, vehicles in dense area are clustered based on their geographical locations, predicted location of vehicles, and speeds. Then UAVs are deployed to serve the clusters. In the second stage, NOMA groups are formed within each cluster and radio resources are allocated to vehicles based on NOMA groups. An optimization problem is formulated and a suboptimal method is used to solve it. The performance of the proposed method is evaluated through simulations where results demonstrate superiority of proposed method in spectral efficiency, min point, and distance.

*Keywords*— Clustering, Radio Resource Management, C-V2X, NOMA, UAV Introduction (Heading 1)

## I. INTRODUCTION

In recent years, there has been a significant increase in the use of smart devices, high data rate applications, and machine-to-machine communications. According to forecasts, by 2030, the number of connected wireless devices, including smartphones, tablets, wearable devices, and sensors, will reach 97.2 billion, generating more than 5036 exabytes of data per month [1]. Accordingly, the new generation of wireless networks must increase their capacity to meet these needs. In the 6th generation (6G) wireless networks, terrestrial, aerial, space, and maritime wireless networks will be integrated into a single network to fulfill the needs of a fully connected global network. Today, the issue of safety in transportation systems is raised as one of the important challenges. Also, the congestion of streets and roads is a problem that is exacerbated by the increasing number of vehicles and urbanization. Vehicle-to-Everything (V2X) communication is a technology that allows vehicles to communicate with their neighboring vehicles, traffic elements, pedestrians, cloud, etc. using wireless signals [2]. The need for fast connectivity, very low latency, and high data transfer rate and reliability makes C-V2X communications challenging [3]. Another notable challenge in vehicular environments is the uneven distribution of vehicular users in some areas/cells. For example, the number of users at intersections and high-traffic roads is very different from that on quiet streets. In fact, neither users in crowded environments receive adequate service quality nor the radio resources are optimally used in quiet environments [4]. To improve performance, advanced multiple access technologies are considered as promising solutions. Non-orthogonal multiple access (NOMA) is studied as a solution in vehicular networks to improve spectrum efficiency [5]. In power domain NOMA, by using power variations on the transmitter side, a transmitter can simultaneously send different traffic information to multiple vehicular terminals on one channel, which improves spectrum efficiency and reduces traffic congestion caused by widespread access [6]. In other words, the transmitter side sends a composite signal consisting of two different signals in the same time and frequency to the receiver side. The two different interfering signals can be decoded by the receiver side due to distinct power levels allocated for transmission to each user. In fact, lower power is used to transmit to the user with a better channel condition, and a signal with higher power is used to transmit to the user with the worse channel condition. Successive Interference Cancellation (SIC) is performed in the node with better channel condition to remove the stronger signal sent to the user with worse channel condition, to decode and recognize the desired signal [8]. To overcome the shortcomings of traditional terrestrial wireless networks, such as low coverage and high deployment costs, Unmanned Aerial Vehicles (UAVs) are becoming a component of the next generation cellular networks. Specifically, by leveraging the flexibility and high mobility of drones, the performance of communication systems can be improved by bringing Drone Base Stations (DBSs) closer to the intended users. In practice, compared to conventional terrestrial communication systems, there is a higher probability of establishing a strong direct line-of-sight connection between DBSs and ground terminals. Also, one of the best solutions for managing and resolving the problem of uneven user congestion in cells is the use of DBSs temporarily in dense areas.

Literature of V2X communications includes both OMA-based [10] and NOMA-based [11] methods. References [9], [12] and [14] consider V2X communications using NOMA, and reference [13] considers NOMA for V2X communications employing UAVs in a single cell network. In reference [9], the resource allocation problem aims to maximize energy efficiency in C-V2X with the help of NOMA in the downlink. Reference [10] defines a resource allocation problem aiming to maximize the data rate in downlink considering the minimum required QoS of vehicles. In reference [11], the downlink resource allocation problem aims to minimize delay in C-V2X with the help of NOMA. This work considers movement of users and dynamic channel information. Reference [12] addresses resource allocation problem to minimize the update cycle time of users' location information. The work uses clustering of vehicles and resource allocation in each cluster. Reference [13] describes a cellular network considering NOMA for Vehicle-to-Vehicle (V2V) and Vehicle-to-Infrastructure (V2I) communications. Authors of [13] allocate resources with the aim of maximizing the data rate and reducing communication delay in single cell C-V2X. This work compares the use of aerial and fixed stations considering delay-sensitive users and delay-tolerant users. However, a single cell scenario has been considered in this method. Reference [14] considers a downlink resource allocation problem aiming to maximize the data rate of users in a two-tier macrocell and femtocell network using vehicle clustering. This paper considers NOMA for V2V communications and V2I in each cluster.

To the best of our knowledge, previous works have not addressed the issue of NOMA-based resource management in multi-cell DBS-assisted C-V2X. Therefore, this paper considers the issue of radio resource management in the V2X network assisted with DBSs and NOMA aiming at maximizing the spectral efficiency of vehicular users in Sub-6GHz downlink communication. Also, in this paper, a clustering method based on the density of vehicles, radius of DBS coverage, speed range of vehicles and predicting the path of vehicles is proposed to indicate the jam areas that need to be supported by DBS. The location of DBS is adjusted based on the clusters' centers. Then, we use a heuristic method to allocate the frequency channels and transmit power.

The rest of the paper is organized as follows. Section 2 is dedicated to describing the system model and formulated problem. Section 3 discusses the proposed method. Section 4 examines the simulation results and Section 5 concludes the paper.

## II. SYSTEM MODEL AND PROBLEM FORMULATION

As seen in Fig. 1, a two-tier vehicular network with V2I communications is considered where vehicles are served by a macrocell and several DBSs. In this network, the spectrum band used by base stations is Sub-6GHz in both Macro Base Station (MBS) and DBSs. To improve the spectrum efficiency for V2X communications, we employ NOMA techniques in downlink V2I communications. The number of vehicles is denoted by $U$. The MBS is denoted by $B$ and has a coverage radius of $r_B$, and the set of DBSs is denoted by $\mathcal{K}$ where each DBS, $k \in \mathcal{K}$, has a coverage radius of $r_k$ and its location is denoted by $d_k(x_k, y_k, z_k)$. The locations of DBSs are considered as one of the variables of our stated problem. The number of vehicles receiving service from macrocell $B$ is denoted by $M$, and the set of macrocell users is denoted by $\mathcal{M}$. The number of users served by DBS $k$ is denoted by $F_k$, and the set of these users is denoted by $\mathcal{F}_k$. It is worth mentioning that each vehicle is served by only MBS or one of the DBSs at any moment. The distribution of vehicles and their movement paths are determined randomly. Each vehicle selects a random destination and travels toward it at a randomly chosen speed within the range of 20 to 50 km/h.

The number of available channels in our NOMA-based system is denoted by $N$, and the set of channels is denoted by $\mathcal{N}$. The binary variable $a_{i,j}^n$ denotes the allocation of channel $n$ to base station $i$ for serving user $j$. If channel $n$ is allocated to base station $i$ and user $j$, $a_{i,j}^n$ is equal to 1; otherwise, it is 0.

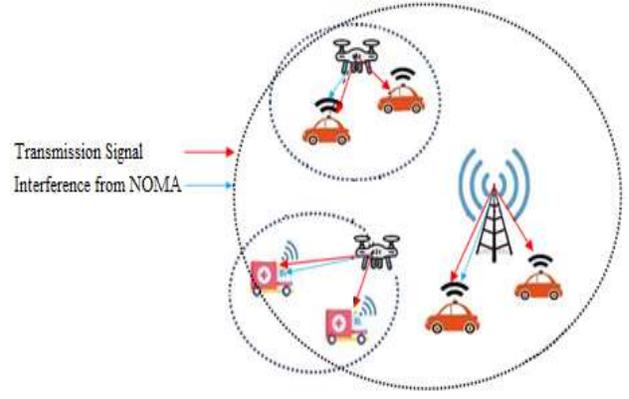

Fig. 1. System Model

Since in NOMA, a channel is usually allocated to a pair of users, user grouping is performed, and then the channels are allocated. The symbol $p_{i,j}^n$ represents the downlink transmission power between base station $i$ (MBS or DBS) and user $j$ on channel $n$. Also, the symbol $h_{i,j}^n$ represents the path gain between transmitter $i$ and receiver $j$ on channel $n$ ($j \in \{\cup_{k=1}^{K} \mathcal{F}_k \cup \mathcal{M}\}, i \in \{B \cup \mathcal{K}\}$). $P_i^{max}$ is the maximum downlink transmission power of base station i ($i \in \{B \cup \mathcal{K}\}$).

In both MBS and DBS, the interference imposed by the NOMA user with better path gain on the user with less path gain is considered. Due to assuming orthogonal channels for MBS users and DBS users, interference between MBS users and DBS users has been disregarded. $R_{i,j}^n$ and $I_{i,j}^n$ respectively represent the maximum achievable spectral efficiency and the amount of interference on the link between base station $i$ and user $j$ on channel $n$. Equations (1) and (2) represent the spectral efficiency for the far user and near user respectively where $\alpha_f$ and $\alpha_n$ represent the power coefficient for the far and near user.

$$R_{i,j}^n = \log_2\left(1 + \frac{\alpha_f \ a_{i,j}^n p_{i,j}^n h_{i,j}^n}{\alpha_n I_{i,j}^n + \sigma^2}\right),$$

$$I_{i,j}^n = p_{i,y}^n \ h_{i,j}^n, i \text{ and } y \text{ are near and far users} \quad (1)$$

$$R_{i,j}^n = \log_2\left(1 + \frac{\alpha_n a_{i,j}^n p_{i,j}^n h_{i,j}^n}{\alpha_f I_{i,j}^n + \sigma^2}\right),$$

$$I_{i,j}^n = 0, \quad j \text{ is far user} \quad (2)$$

Another challenge in power allocation is adhering to the SIC constraint. As seen in Equation (3), the difference in the SINR of two users of a pair must be greater than the minimum threshold required for successful SIC, denoted by $P_{tol}$.

$$\frac{a_{i,j}^n p_{i,j}^n h_{i,j}^n}{I_{i,j}^n + \sigma^2} - \frac{a_{i,y}^n p_{i,y}^n h_{i,y}^n}{I_{i,y}^n + \sigma^2}$$
$$\geq P_{tol} \ if \ a_{i,j}^n = a_{i,y}^n = 1, i$$
$$\in \{B \cup \mathcal{K}\}, j, y$$
$$\in \{\cup_{k=1}^{K} \mathcal{F}_k \cup \mathcal{M}\}$$

(3)

$$\max_{\{a_{k,f}^n, p_{k,f}^n, a_{B,m}^n, p_{B,m}^n, d_k\}} \sum_{K=1}^{K}\sum_{f=1}^{F_k}\sum_{n=1}^{N} a_{k,f}^n \log_2\left(1+\frac{p_{k,f}^n h_{k,f}^n}{I_{k,f}^n+\sigma^2}\right) + \sum_{m=1}^{M}\sum_{n=1}^{N} a_{B,m}^n \log_2\left(1+\frac{p_{B,m}^n h_{B,m}^n}{I_{B,m}^n+\sigma^2}\right)$$

subject to

$$C1: \frac{a_{i,j}^n p_{i,j}^n h_{i,j}^n}{I_{i,j}^n+\sigma^2} - \frac{a_{i,y}^n p_{i,y}^n h_{i,y}^n}{I_{i,y}^n+\sigma^2} \geq P_{tol},\ a_{i,j}^n = a_{i,y}^n = 1, \forall i \in \{B \cup K\}, j, y \in \left\{\bigcup_{k=1}^{K} F_k \cup M\right\}, n \in N$$

$$C2: \sum_{f=1}^{F_k}\sum_{n=1}^{N} p_{k,f}^n \leq P_k^{\max}, \forall k \in K$$

$$C3: \sum_{m=1}^{M}\sum_{n=1}^{N} p_{B,m}^n \leq P_B^{\max}$$

$$C4: \sum_{f=1}^{F_k} a_{k,f}^n \leq 2, \forall k \in K, n \in N$$

$$C5: \sum_{m=1}^{M} a_{B,m}^n \leq 2,\ \forall n \in N$$

$$C6: a_{i,j}^n \in \{0,1\},\ p_{i,j}^n \geq 0, \forall i \in \{K \cup B\}, j \in \{F_k \cup M\}, n \in N \tag{5}$$

The total spectral efficiency of macrocell and DBS users, denoted by $R$, is shown in Equation (4).

$$R = \sum_{k=1}^{K}\sum_{f=1}^{F_k}\sum_{n=1}^{N} a_{k,f}^n R_{k,f}^n + \sum_{m=1}^{M}\sum_{n=1}^{N} a_{B,m}^n R_{B,m}^n \tag{4}$$

We employ user clustering, DBS deployment, power allocation, and radio resource allocation, to achieve this goal. The formulated optimization problem is shown in Equation (5).

In Equation (5), C1 refers to the adherence to the SIC-related constraint. Constraints C2 and C3 ensure that the sum of the transmission power of DBSs and MBS is less than their maximum values respectively. Constraints C4 and C5 respectively indicate that each channel is allocated to only one group of vehicular users connected to DBSs and MBS. Constraint C6 also refers to the binary nature of channel allocation and the non-negativity of the transmission power of DBSs and MBS.

III. PROPOSED METHOD

Generally, due to the non-convexity and the discrete nature of the channel allocation, the problem of Equation (5) is a non-convex MINLP. Based on the decomposition method, solving this problem occurs in 3 stages. First, we proceed to cluster the vehicles in dense area and deploy DBSs in appropriate locations to serve them. Vehicles that are not included in any clusters receive service from the MBS. Then, we pair the vehicles in each cluster to form NOMA groups. At this stage, we solve a channel and power allocation sub-problem for MBS and DBSs users.

A. Clustering vehicles and deploying DBSs

For the clustering, we use a custom version of DBSCAN algorithm. DBSCAN algorithm is exploited as a density-based clustering method to make clusters in high-density congregation of vehicular users. In this method, two parameters, i.e., the maximum distance between two users ($\varepsilon$) and the minimum required members (*MinPoints*), should be adjusted. This algorithm first selects a member (which is a vehicle) and searches for its neighbors based on the distance between the two users. A cluster is formed if at least *MinPoints* vehicles are chosen as its members. In our improved DBSCAN method, in addition to the *MinPoints* and distance parameters, the speed and direction of the vehicles are considered.

Initially, we consider a vehicle as the first member of a candidate cluster. Now, based on the distance parameter, we determine the neighbors of this vehicle. Then, we determine the center of the formed cluster and calculate the distance of each member to it. The *Dis(i, center)* denotes the distance of vehicle i to the center. If any of these neighbors is outside the range of the DBS which is assumed in center of the candidate cluster or is out of the range of speed with the other member of the cluster or their predicted location of vehicles is different, they will be removed from the neighborhood. *SR(i, j)* and *PL(i, j)* respectively represent the vehicles *i* and *j* have speed difference lower than *Θ* and their predicted future locations are the same. For location prediction, we assume method of [16]. Now, if the number of neighboring members of this vehicle is more than *MinPoints*, we will create a cluster. Then, to expand the cluster, we repeat all the steps for each of the neighboring vehicles. If by adding a vehicle, more than one vehicle goes out of the neighborhood list, we will not add that vehicle. We continue these steps to check all vehicles. Finally, vehicles not assigned to any cluster will receive service from MBS. The DBSCAN-based vehicle clustering is presented in Algorithm 1.

| | **Algorithm 1**: DBSCAN-based vehicle clustering |
|---|---|
| | **Input:** |
| | *X:* set of all vehicles |
| | *Radius*: DBS coverage radius |
| | **Begin** |
| 1 | **For** *i*=1 To |*X*| |
| 2 | *Ngb(X[i])* = All neighbors of *X[i]* regarding distance threshold (*ε*) |
| 3 | *Center* = Calculate the center of the cluster formed by *Ngb (X[i])* |
| 4 | **For** *j* = 1 To |*Ngb(X[i])*| |
| 5 | **If** (*Dis(Ngb(X[i])[j], Center) > Radius* ‖ !*SR(X[i],Ngb(X[i])[j])* ‖ !*PL(X[i], Ngb(X[i])[j])*) |
| 6 | remove *j* from *Ngb(X[i])* |
| 7 | **End If** |
| 8 | **End For** |
| 9 | **If** (|*Ngb(X[i])*|>= *MinPoints*) |
| 10 | Create a cluster including *i* and *Ngb(X[i])* |
| 11 | **End If** |
| 12 | Expand created cluster repeating the algorithm per every vehicle in *Ngb(X[i])* |
| 13 | **End For** |

Using Algorithm 1, we obtain several clusters whose vehicles are in a vicinity and similar to each other in terms of speed and regarding the future locations.

After forming the clusters, we go on to determine the position of each DBS that serves a cluster. At this stage, the DBS placement is done based on the average position of the vehicles inside each cluster and its position changes as the mean of the cluster is changed, as the above stages are executed iteratively. Our clustering method leads the DBSs serve the vehicles of clusters for a longer time as the movement and proximity of vehicles are similar in a cluster.

### B. Pairing devices and allocating resources

After determining the location of the DBSs, we proceed to form pairs of users inside each cluster. Clusters use orthogonal resources to avoid inter-cluster interference. In this stage, we first sort the users based on the path gain to the DBS and then proceed to form pairs of vehicles, far and near. For example, if we have 10 vehicles in a cluster, after sorting them based on the path gain, i.e., vehicle 1 to vehicle 10, we pair vehicle 1 with vehicle 6, vehicle 2 with vehicle 7, and so on. If a user does not join any NOMA pair, it receives service in OMA mode.

After forming NOMA pairs, the stated problem turns into a sub-problem of channel allocation and transmission power assignment. This sub-problem is like Equation (5) but assuming the location of DBSs. To solve this sub-problem, we use decomposition method and execute the following sub-problems iteratively. First, we allocate the channels solving a sub-problem and then we adjust the transmission power.

For channel allocation sub-problem, we assume a fixed transmit power for each user. Now, according to the objective function defined in Equation (5), which is to maximize the total spectral efficiency, and based on Equation (1), there is a direct relationship between the path gain and the spectral efficiency. Accordingly, to maximize the total spectral efficiency, channels must be allocated in a way to maximize the total path gain of the vehicles. For this purpose, we formulate the channel allocation sub-problem as in Equation (6).

$$\max_{\{a_{k,f}^n, a_{B,m}^n\}} \sum_{K=1}^{K}\sum_{f=1}^{F_k}\sum_{n=1}^{N} a_{k,f}^n h_{k,f}^n + \sum_{m=1}^{M}\sum_{n=1}^{N} a_{B,m}^n h_{B,m}^n$$

subject to

C1: $\sum_{f=1}^{F_k} a_{k,f}^n \leq 2, \forall k \in \mathcal{K}, n \in \mathbb{N}$

C2: $\sum_{m=1}^{M} a_{B,m}^n \leq 2, \forall n \in \mathbb{N}$

C3: $a_{k,f}^n, a_{B,m}^n, \forall i \in \{\mathcal{K} \cup B\}, j \in \{F_k \cup M\}, n \in \mathbb{N}$ (6)

where its objective function is to maximize the sum of the path gains. C1 and C2 restrict the number of allocated channels to each DBS and MBS user. Finally, C3 indicates that channel allocation is a binary variable.

The optimization problem of Equation (6) is an assignment problem which can be solved using the Hungarian method [15]. To solve it, we assume channels as jobs, MBS and DBS users as employees, and $-h_{i,j}^n$ as the cost of assigning channel $n$ to user $j$ of base station $i$ ($j \in \{\bigcup_{k=1}^{K} \mathcal{F}_k \cup \mathcal{M}\}, i \in \{B \cup \mathcal{K}\}$).

Finally, to solve power allocation sub-problem, we use a heuristic method. The power allocated to all NOMA pairs is considered equal. Accordingly, the power of the far and near users in each NOMA pair is calculated based on Equations (7) and (8), respectively. The $nc$ represents the number of groups in every cluster.

$$p_{i,j}^n = \alpha_f \frac{P_i^{max}}{nc}, j \text{ is near user} \in \left\{\bigcup_{k=1}^{K} \mathcal{F}_k \cup \mathcal{M}\right\}, i \in \{B \cup \mathcal{K}\} \quad (7)$$

$$p_{i,y}^n = \alpha_n \frac{P_i^{max}}{nc}, y \text{ is far user} \in \left\{\bigcup_{k=1}^{K} \mathcal{F}_k \cup \mathcal{M}\right\}, i \in \{B \cup \mathcal{K}\} \quad (8)$$

### IV. SIMULATION RESULTS

The proposed method is simulated using a custom simulator in MATLAB. In this section, we first describe the simulation settings. Then, we compare our proposed method with the method in [14]. In [14], NOMA is used for V2V and V2I communications. The method of [14] aims to maximize the total spectral efficiency of vehicular users connected to macrocell and femtocell. The total spectral efficiency of users is compared based on the number of vehicles, the clustering distance criterion, and the minimum point clustering criterion.

To evaluate the proposed method, the evaluation criterion is the total spectral efficiency of vehicular users. The path-loss between the base station i and the vehicle j is modeled as $PL_{i,j}$=28.1 + 37.6 * $\log_{10}(d(i,j))$+$\log_{10}(fc/2.5)$, where $d(i,j)$ is the distance between the base station i and the user j and fc is the carrier frequency. The results are the average of 500 runs with independent random seeds. The parameters used in the simulation are stated in Table 1. Fig. 2 shows the total spectral efficiency of all vehicles versus the DBS coverage radius, where *ε*=4, *MinPoints*=6 and the DBS coverage radius increases from 5 to 14 m for our proposed algorithm and method of [14]. As can be seen, with the increase in the coverage radius of the DBS, we initially face an increase and then a decrease in the total spectral efficiency.

TABLE I.  SIMULATION PARAMETER

| Parameter | Value |
|---|---|
| $P_B^{max}$ | 20 w |
| $P_K^{max}$ | 2 w |
| $U$ | 50 |
| $N_0$ | -80 dBm/Hz |
| $r_B$ | 500 m |
| $r_K$ | 10 m |
| $f_c$ | 2.5 GHz |

The reason is that as the coverage radius increases, the number of clusters decreases, and more vehicles are placed in each cluster. This causes the number of NOMA groups to be increased and less transmission power to be allocated to each NOMA group. As a result, the total spectral efficiency decreases. Also, it can be seen that our proposed algorithm has better performance than the method of [14] in terms of total date rate.

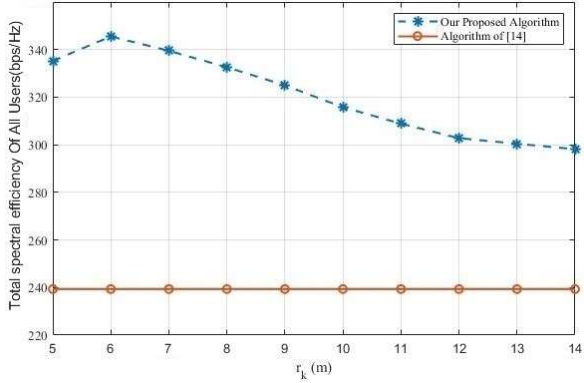

Fig. 2. *Total Spectral efficiency versus DBS coverage radius*

In Fig. 3, the changes in the spectral efficiency of vehicles versus the increase of $\varepsilon$ for *MinPoints* =6 and DBS *coverage radius*=10 m are shown. As can be seen, with the addition of the $\varepsilon$, the total spectral efficiency of users first increases and then remained almost constant. In fact, initially, as $\varepsilon$ increases, the number of vehicles that can receive service from DBS will be increased. This issue leads to an increase in the total spectral efficiency. However, the coverage radius of the DBS allows for an increase in $\varepsilon$ up to a certain point, after which it causes the total spectral efficiency to remain constant. Also, it can be seen that our proposed algorithm has better performance than the algorithm presented in [14] in terms of aggregate date rate, when the $\varepsilon$ is increased.

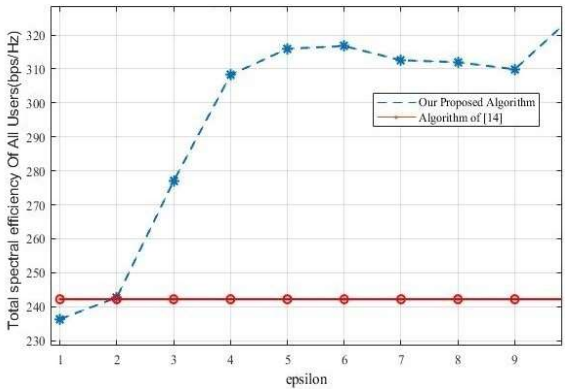

Fig. 3. *Total Spectral efficiency versus clustering distance, ε*

Fig. 4 shows the total spectral efficiency of all vehicles versus *MinPoints*, where $\varepsilon$=4, the DBS coverage radius is 10 and MinPoints increases from 4 to 13 for our proposed algorithm. We can see that the total spectral efficiency of vehicles decreases as the MinPoints increases. The reason is that as the MinPoints increases, the number of clusters decreases, and more vehicles are placed in each cluster. This causes the number of NOMA groups to increase and less transmission power to be allocated to each NOMA group. As a result, the total spectral efficiency decreases. Also, it can be seen that our proposed algorithm has better performance than the algorithm presented in [14] in terms of total spectral efficiency.

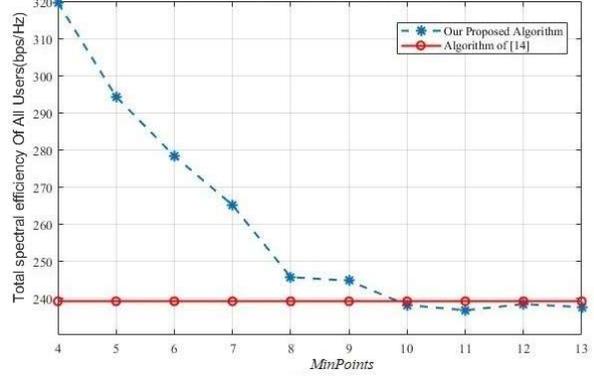

Fig. 4. *Total Spectral efficiency versus the MinPoints value*

## V. CONCLUSION

In this paper, radio resource management in a V2X network assisted with DBSs and NOMA was investigated aiming at maximizing the spectral efficiency of vehicular users in Sub-6GHz downlink communication. Given the non-convex nature of the defined problem, it was decomposed into sub-problems. To solve this problem, we first clustered the vehicles using the improved DBSCAN method based on the density of vehicles, radius of DBS coverage, speed range of vehicles and predict location of vehicles. After forming the clusters, we determined the position of each DBS that serves a cluster. Then, in each cluster, NOMA groups were formed based on the path gain of vehicular users. Subsequently, the frequency channel and transmission power were allocated with the aim of maximizing the total spectral efficiency of users. Finally, based on the simulation results, we demonstrated that the proposed method is superior to a previous method in terms of the total spectral efficiency. The simulations show that the coverage radius of the DBS, $\varepsilon$ and *MinPoints* have important effect on the results and should be adjusted appropriately. Improving the performance of our method with considering mmWave frequency spectrum in DBS tier is suggested as future a work. Moreover, considering unequal transmission power between NOMA pairs is considered other suggestion for future work.

## ACKNOWLEDGMENT

This work is supported by National Science Foundation under Grant No. 2219742 and Grant No. 2131001.